\documentclass{article}

\usepackage{arxiv}

\usepackage[utf8]{inputenc} 
\usepackage[T1]{fontenc}    
\usepackage{hyperref}       
\usepackage{url}            
\usepackage{booktabs}       
\usepackage{amsfonts}       
\usepackage{nicefrac}       
\usepackage{microtype}      
\usepackage{lipsum}		
\usepackage{graphicx}
\usepackage{natbib}
\usepackage{doi}
\usepackage{subfig}
\usepackage{amsmath}
\usepackage{authblk}

\title{Performance Calibration of the Wavefront Sensor's EMCCD Detector for the Cool Planets Imaging Coronagraph Aboard CSST}

\author[1,2,*]{Jiangpei Dou}
\author[1,2,*]{Bingli Niu} 
\author[1,2,*]{Gang Zhao}
\author[1,2]{Xi Zhang}
\author[1,2]{Gang Wang}
\author[1,2]{Baoning Yuan}
\author[1,2]{Di Wang}
\author[1,2,3]{Xingguang Qian}

\affil[1]{Nanjing Institute of Astronomical Optics \& Technology, Chinese Academy of Sciences, Nanjing 210042, China}
\affil[2]{CAS Key Laboratory of Astronomical Optics \& Technology, Nanjing Institute of Astronomical Optics \& Technology, Nanjing 210042, China}
\affil[3]{University of Chinese Academy of Sciences, Beijing 100049, China}

\renewcommand{\shorttitle}{Dou et al.}

\hypersetup{
pdftitle={A template for the arxiv style},
pdfsubject={q-bio.NC, q-bio.QM},
pdfauthor={David S.~Hippocampus, Elias D.~Striatum},
pdfkeywords={First keyword, Second keyword, More},
}

\begin{document}
\maketitle

\begin{abstract}
	The wavefront sensor (WFS), equipped with an electron-multiplying charge-coupled device (EMCCD) detector, is a critical component of the Cool Planets Imaging Coronagraph (CPI-C) on the Chinese Space Station Telescope (CSST). Precise calibration of the WFS's EMCCD detector is essential to meet the stringent requirements for high-contrast exoplanet imaging. This study comprehensively characterizes key performance parameters of the detector to ensure its suitability for astronomical observations. Through a multi-stage screening protocol, we identified an EMCCD chip exhibiting high resolution and low noise. The electron-multiplying gain (EM Gain) of the EMCCD was analyzed to determine its impact on signal amplification and noise characteristics, identifying the optimal operational range. Additionally, noise properties such as readout noise were investigated. Experimental results demonstrate that the optimized detector meets CPI-C's initial application requirements, achieving high resolution and low noise. This study provides theoretical and experimental foundations for the use of EMCCD-based WFS in adaptive optics and astronomical observations, ensuring their reliability for advanced space-based imaging applications
\end{abstract}


\section{Introduction}
Exoplanet exploration represents a pivotal scientific endeavor of the 21st century, advancing our understanding of life's origins and humanity's cosmic significance \citep{Seidel2023,galicher2023imaging,SwainMarkR2024}. Primary objectives encompass discovering Earth-sized planets within habitable zones and detecting biosignatures. As an extension of the Copernican revolution, this field may redefine conceptions of life's cosmic centrality. Its significance is underscored by inclusion in the National Research Council's strategic plans and the 2019 Nobel Prize in Physics, awarded for discovering an exoplanet orbiting a solar-type star \citep{MAYORM1995}. Future telescopes and satellites will prioritize this field, reflecting its fundamental scientific importance.

The China Space Station Telescope (CSST), a large space-based observatory, is under development within China's human spaceflight program \citep{Miao2022,Gu2024,Song2024}. With a 2-meter aperture, it combines a wide field of view with high image quality and features on-orbit maintenance and upgrade capabilities \citep{Yao2024,Wen2024,Wang2024}. The Cool Planets Imaging Coronagraph (CPI-C) onboard the CSST provides a critical opportunity for exoplanet exploration, achieving an imaging contrast of up to $10^{-8}$ in the visible and near-infrared spectra, marking a significant milestone in the direct imaging of cold exoplanets \citep{Ren2012,Dou2015,Guo2017,Zhu2021}. Achieving such contrast demands stringent wavefront precision. CPI-C addresses this issue by utilizing a multi-channel electron-multiplying charge-coupled device (EMCCD) in its wavefront sensor. EMCCDs provide three advantages over conventional CCDs: (1) sub-electron readout noise via electron-multiplying gain (EM Gain) \citep{Robbins2003} and single-photon sensitivity \citep{Ottavia2011}, enabling high-precision wavefront sensing under low flux and enhancing adaptive optics efficiency \citep{Zhu2016}; (2) high-speed response suppressing dynamic wavefront distortions from turbulence or jitter  \citep{Tulloch2011}; (3) cryogenic operation with dynamic EM Gain adjustment to suppress dark current and clock-induced charge (CIC) \citep{Harding2016}, maintaining wide dynamic range, linearity, radiation tolerance, and low power consumption \citep{Cagigas2015}. 

EMCCD technology advances high-contrast wavefront metrology. Unlike conventional Shack--Hartmann or pyramid wavefront sensors (SH-WFS/PWFS), which are limited in photon-starved regimes \citep{Burada2017,OyarznF2024}, EMCCDs overcome this via on-chip impact ionization gain. This enables precise microlens spot centroid localization with faint signals, which is critical for space coronagraphy. We advance this technology by systematically calibrating EMCCD noise characteristics and spatial resolution parameters. Experimentally, we confirm the EMCCD detector's capacity to suppress readout noise, meeting CPI-C's stringent wavefront sensing requirements. These requirements are critical for maintaining precision at the \linebreak  $\lambda$/100 level, which is a fundamental prerequisite for achieving the targeted $10^{-8}$ imaging contrast. It is important to note that the performance calibration of the WFS's optical components (e.g., microlens array alignment and wavefront splitting accuracy) will be discussed in detail in a separate publication. This study systematically evaluates the performance of the EMCCD detector across key parameters, including EM Gain, noise characteristics, and non-uniformity across channels, to ensure compliance with mission-critical engineering requirements \citep{Liu2015,Dou2016,Ren2011,Dou2011,BingDong2011}.

Section \ref{sec2} details the detector structure and specifications; Section \ref{sec3} presents calibration results; Section \ref{sec4} discusses findings; and Section \ref{sec5} summarizes conclusions.

\section{Structure and Operating Principle of the WFS's EMCCD Detector}\label{sec2}
\subsection{Structure and Design Specifications}
\begin{figure}
\centering
\includegraphics[width=5 cm]{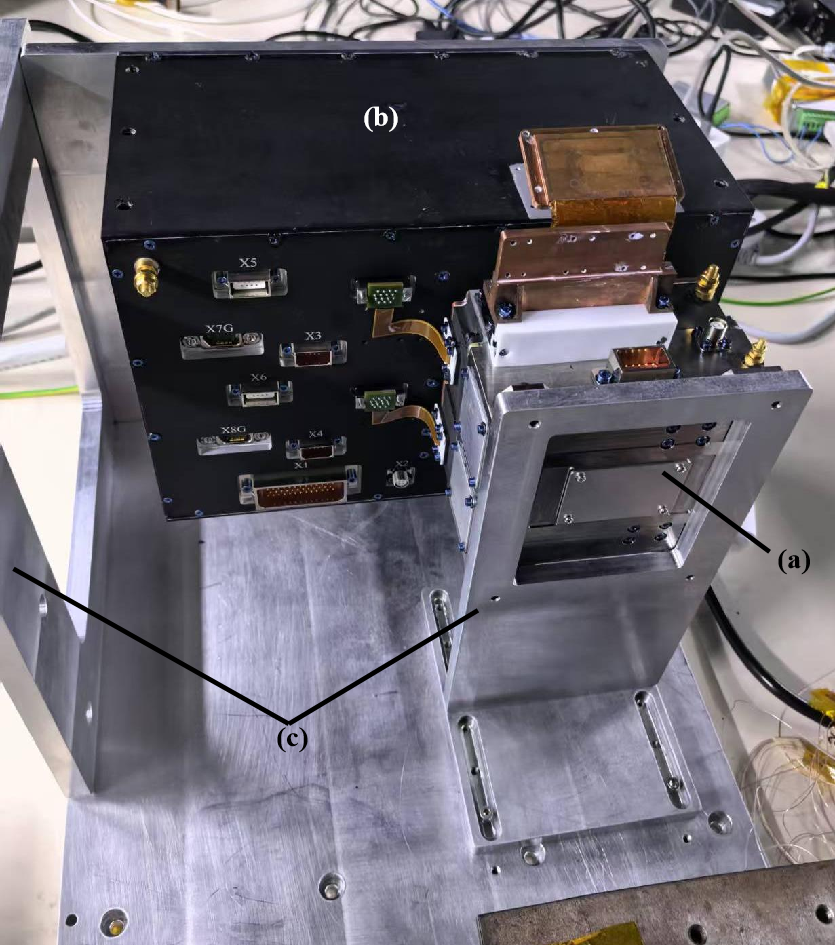}

\caption{Physical configuration of the WFS's EMCCD detector: (a) imaging assembly incorporating the EMCCD chip; (b) electronic control and power supply unit; (c) integrated support structure with heat-pipe-based thermal management.\label{fig1}}
\end{figure}

The detector consists of three core components: (a) an imaging assembly integrating the multi-channel EMCCD chip for optical signal acquisition, (b) an electronic control and power supply unit handling signal processing and power management, and (c) a composite support structure providing mechanical stabilization with integrated heat-pipe thermal management (Figure \ref{fig1}). To achieve CPI-C's high-contrast objectives, we performed a multi-level requirements breakdown spanning top-level contrast metrics, wavefront control accuracy ($\lambda$/100), and detector noise specifications. {This process enabled rigorous quantification of constraints for design parameters} (Table \ref{tab1}), ensuring performance compliance with CPI-C operational requirements. 

\begin{table}
	\caption{Key design specifications of the detector.}
	\centering
		\begin{tabular}{lll}
		\toprule
		\textbf{Specifications}	& \textbf{Value}	\\
		\midrule
		CCD type	& 8-channel EMCCD		\\
		Image readout quantization number & 14 bit			 \\
		Exposure time & 3 ms$_{\textasciitilde}$5 s, adjustment interval 1 ms \\
		EM Gain & Max value $\geq$ 100$\times$ \\
		
		\bottomrule
	\end{tabular}
	\label{tab1}
\end{table}

\subsection{Chip Screening and Selection Process}
\begin{figure}

\centering
\includegraphics[width=7 cm]{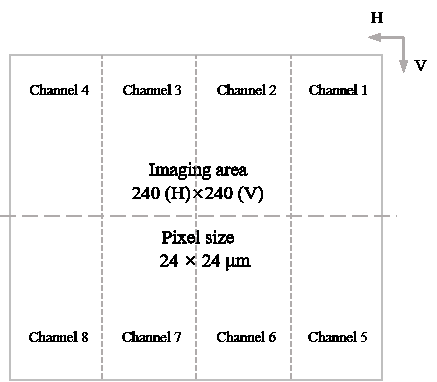}

\caption{EMCCD focal plane schematic: Storage section with 8 parallel readout channels, each channel: 60(H) $\times$ 120(V) pixels, and output amplifiers shared pairwise (channels 1-2, 3-4, 5-6, 7-8).\label{fig2}}
\end{figure}

The EMCCD chip, which is central to the operation of the WFS's EMCCD detector, delivers high sensitivity, low noise, and wide dynamic range, as shown in Figure \ref{fig2}. To meet CPI-C requirements, we implemented a multi-stage chip screening protocol. During development, chips underwent sequential evaluation via a standardized test circuit. These evaluations used uncalibrated conditions to objectively characterize inherent process variations. Selection criteria included two key metrics: image noise (lower values indicate superior signal integrity) and image sharpness (quantified via edge contrast analysis). This dual-metric framework ensured only chips excelling in both noise control and resolution advanced to integration.
\vspace{-3pt}

\subsubsection{Image Noise Comparison via Standard Deviation}
\begin{figure}

\subfloat[\centering]{\includegraphics[width=6.5cm]{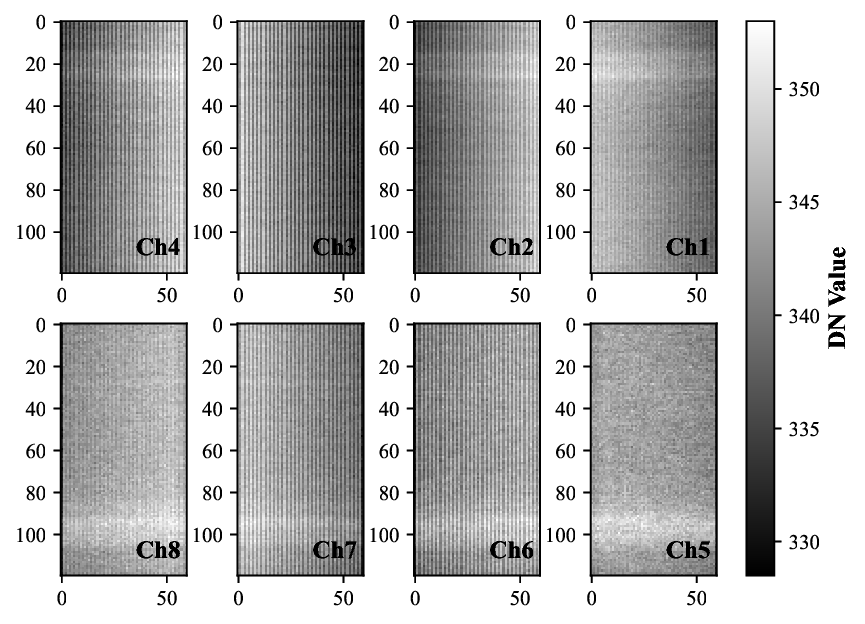}}
\hfill
\subfloat[\centering]{\includegraphics[width=6.5cm]{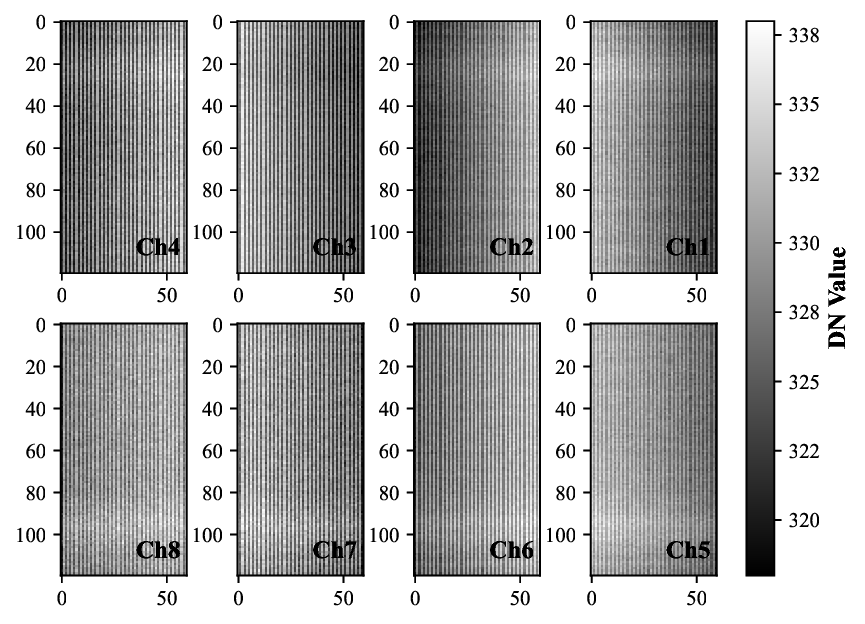}} \\
\hspace{5cm}
\subfloat[\centering]{\includegraphics[width=6.5cm]{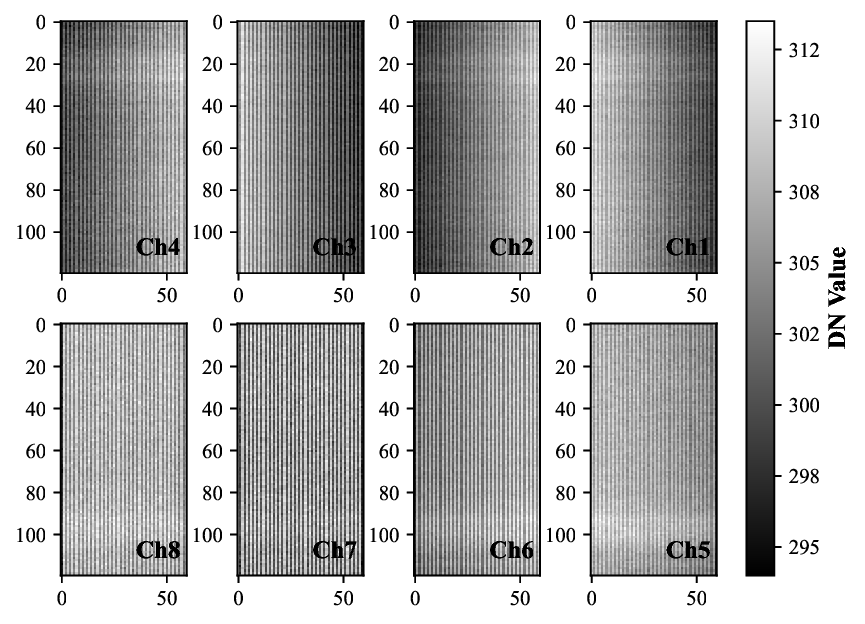}}
\hfill
\subfloat[\centering]{\includegraphics[width=6.5cm]{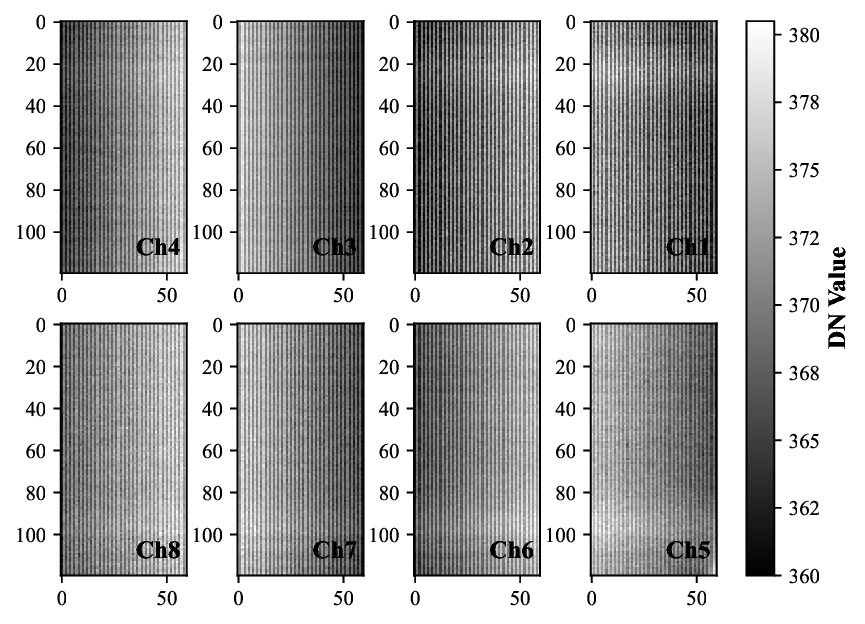}}\\

\caption{{Bias} 
 frame characteristics of candidate EMCCD chips (with 8 channels (Ch) per chip):\linebreak   (\textbf{a}) Chip 1, (\textbf{b}) Chip 2, (\textbf{c}) Chip 3, (\textbf{d}) Chip 4.\label{fig3}}

\end{figure}

Image noise, quantified by the digital number (DN) standard deviation ($\sigma$), is a critical metric for evaluating signal fidelity, with lower $\sigma$ values reflecting superior noise performance. We standardized testing conditions for all EMCCD chips: EM Gain disabled, exposure time minimized to 3 ms, and 100-frame averaging applied. Representative bias frames acquired under these standardized conditions are presented in Figure\ref{fig3}.

\begin{figure}
\centering
\includegraphics[width=11.5 cm]{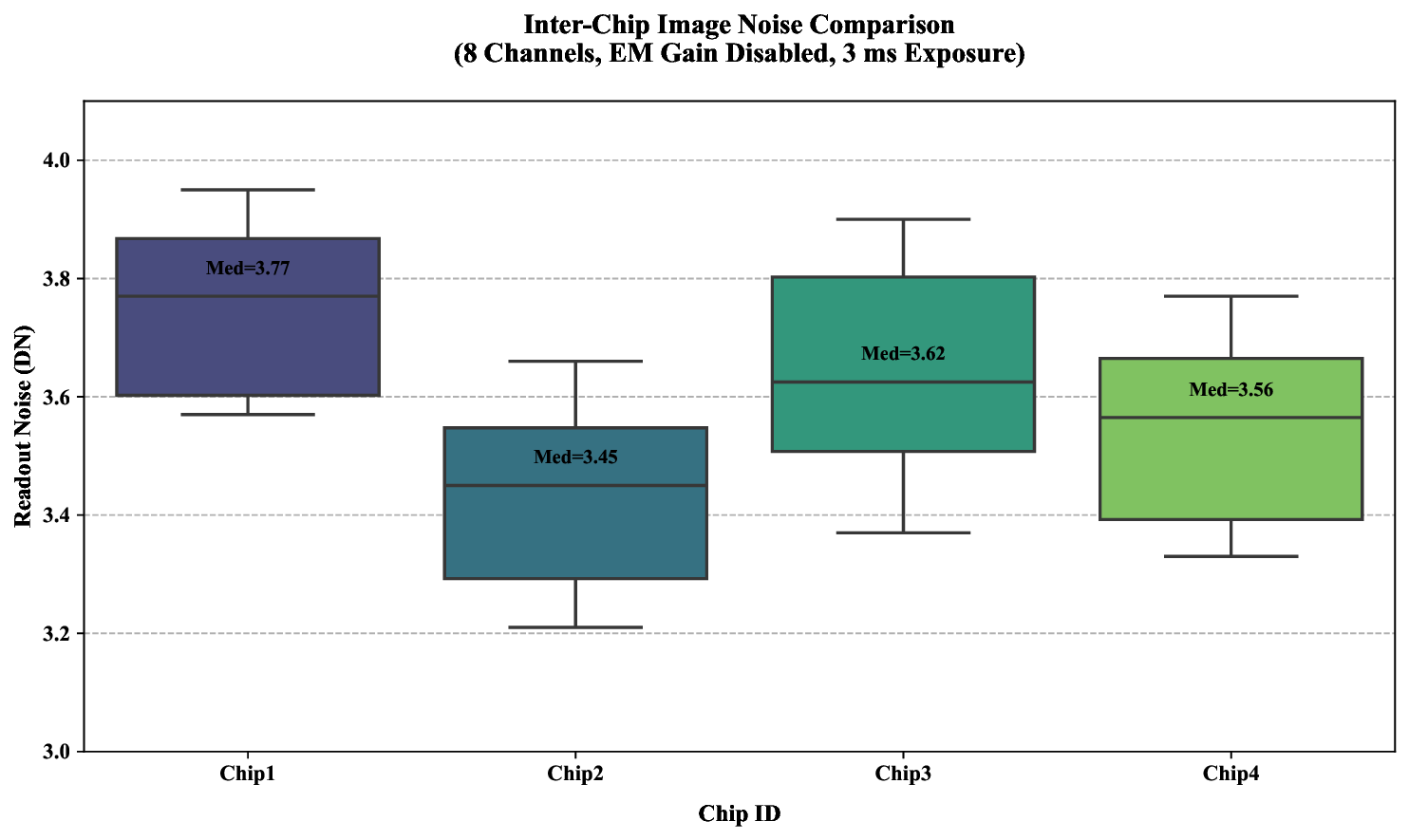}

\caption{Image noise characterization and inter-chip comparison.\label{fig4}}
\end{figure}

Figure \ref{fig4} shows significant inter-chip noise variations via boxplot analysis. Chip~2 exhibited optimal performance with the lowest median readout noise (3.45 DN) and consistent channels. Conversely, Chip 1 showed the highest median noise (3.77 DN) despite similar channel stability. 
\vspace{-6pt}

\unskip

\subsubsection{Image Sharpness}
\begin{figure}

\centering
\subfloat[\centering]{\includegraphics[width=4cm]{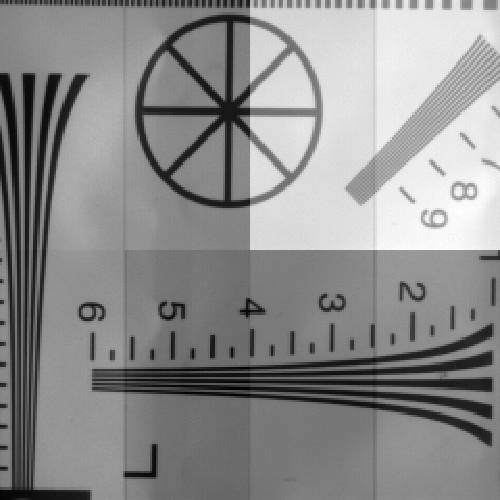}}
\hfill
\subfloat[\centering]{\includegraphics[width=4cm]{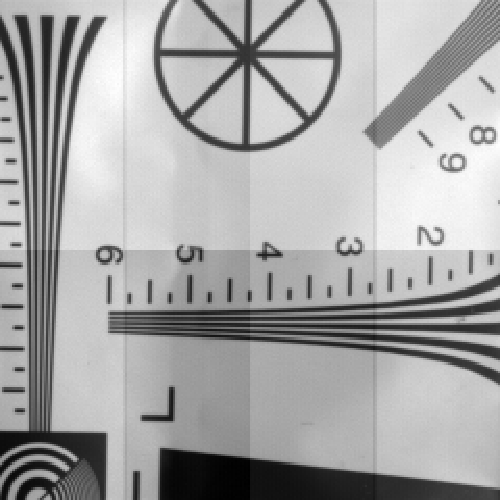}}
\hfill
\subfloat[\centering]{\includegraphics[width=4cm]{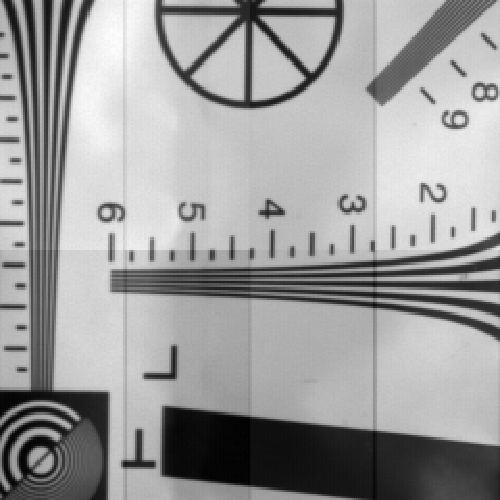}}
\hfill
\subfloat[\centering]{\includegraphics[width=4cm]{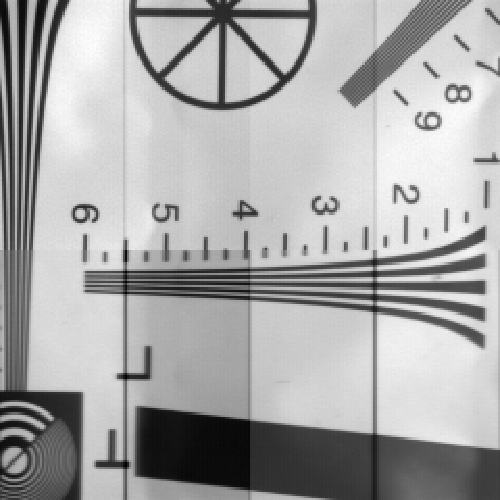}}\\

\caption{(\textbf{a}) Chip 1 with Laplacian variance of 1413, demonstrating moderate image sharpness. (\textbf{b}) Chip 2 with Laplacian variance of 1832, exhibiting enhanced edge contrast and superior clarity. \mbox{(\textbf{c}) Chip 3} with Laplacian variance of 1319, presenting balanced noise suppression and resolution. \mbox{(\textbf{d}) Chip} 4 with Laplacian variance of 2165, achieving the highest level of detail retention and texture definition.The Laplace variance exhibits a positive correlation with resolution. Lower values suggest superior noise performance, whereas higher values indicate finer resolution capabilities.\label{fig5}}
\end{figure}

{Image sharpness assesses clarity and resolution, with higher values being critical for accurate wavefront reconstruction. We quantified sharpness by applying a Laplacian operator and calculating its variance} \citep{Bansal2016BlurID}, {where this variance serves as an image clarity metric.} Figure \ref{fig5} presents the sharpness performance of the tested EMCCD chips.

Based on selection results, Chip 2 and Chip 4 were designated for the flight model (FM) and qualification model (QM) units, respectively. The system is currently in QM phase, with subsequent performance calibrations analyzed using QM unit test data.

\section{Calibration Results of Key Performance Parameters for the Detector}\label{sec3}
\begin{figure}
\centering
\includegraphics[width=10 cm]{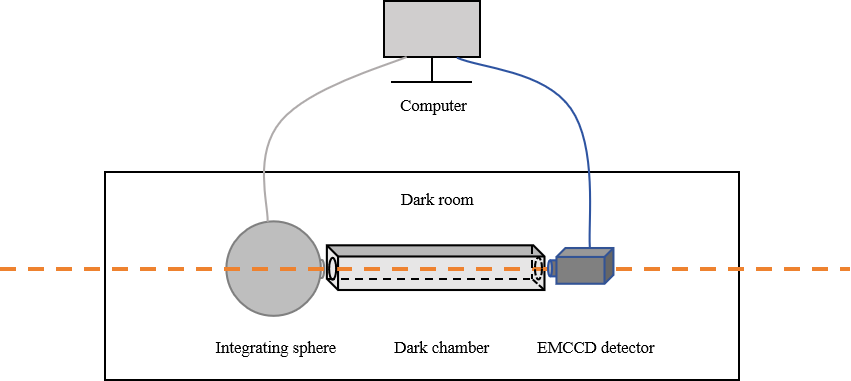}

\caption{Schematic diagram of the test system.}\label{fig6}
\end{figure}

We evaluated key detector performance characteristics including EM Gain \citep{Tulloch2011,RyanDuncanP2021}, readout noise \citep{Niu2022}, and channel non-uniformity. This assessment validates sensor reliability and accuracy for scientific and engineering applications. Experiments were conducted in a strictly controlled darkroom. A standard uniform light field was generated using an integrating sphere light source, which was combined with a customized dark chamber to shield against stray light in the surroundings. Consequently, an optical path system was established (Figure \ref{fig6}). Prior to experiments, we verified photometric flatness of the integrating sphere and dark chamber outlet to ensure compliance with the uniformity requirements for planar calibration light sources. An industrial control computer managed the experimental process. Following one hour of cooling and stabilization for the detector, operations are executed sequentially: (1) The integrating sphere light source is turned off to capture the bias images. (2) The intensity of the light source is dynamically adjusted within the unsaturated range of the detector while maintaining a constant output. \mbox{(3) By interlocking} EM Gain and exposure time adjustments of the detector, {we acquired} a comprehensive set of multi-parameter combined flat-field calibration images is ultimately obtained, providing a high-precision reference for subsequent data calibration.

\subsection{EM Gain}\label{sec31}

The detector amplifies weak signals via voltage-controlled electron multiplication, with EM Gain precisely regulated through injection voltage adjustments. EM Gain exhibits significant temperature sensitivity, {characterized by exponentially amplified nonlinear responses} as the cooling temperature decreases \citep{Harding2016}. To ensure an optimal balance between system stability and operational efficiency of the detector, this study conducted systematic calibration of the coupling relationship between EM Gain and cooling temperature under laboratory conditions and safety specifications. Experimental validation {was performed} across three temperature regimes ($-$15 \textdegree C, $-$20 \textdegree C, $-$25 \textdegree C). This calibration aims to \linebreak  (1) establish a reliable working boundary parameter system by quantifying the nonlinear dependence of gain on temperature, and (2) analyze how EM Gain varies with cooling temperature to determine optimal control conditions that meet specified gain requirements. The EM Gain calibration was implemented through the following formula:

\begin{equation}
\text{EM Gain} = \frac{\text{flat}_{\text{Gm}} - \text{bias}_{\text{Gm}}}{\text{flat}_{\text{1}} - \text{bias}_{\text{1}}} \times \frac{\text{exptime}_{\text{1}}}{\text{exptime}_{\text{Gm}}}
\end{equation}

\begin{figure}

\centering
\subfloat[\centering]{\includegraphics[width=6cm]{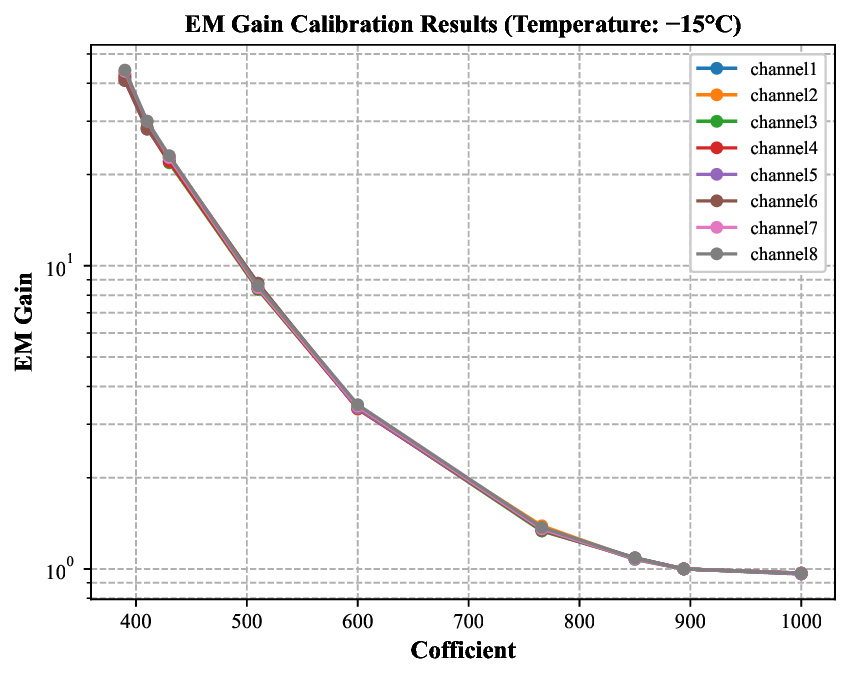}}
\subfloat[\centering]{\includegraphics[width=6cm]{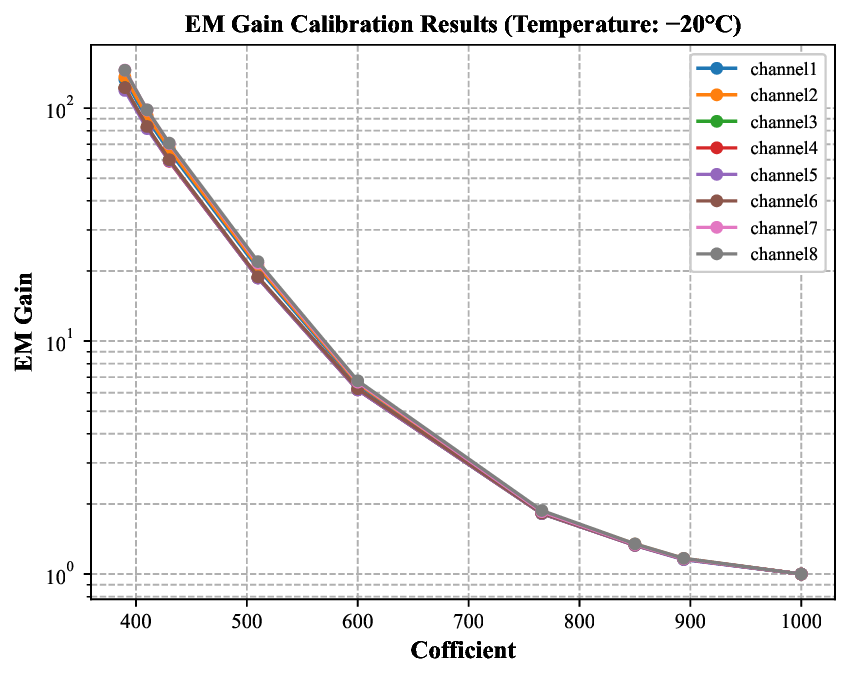}}
\subfloat[\centering]{\includegraphics[width=6cm]{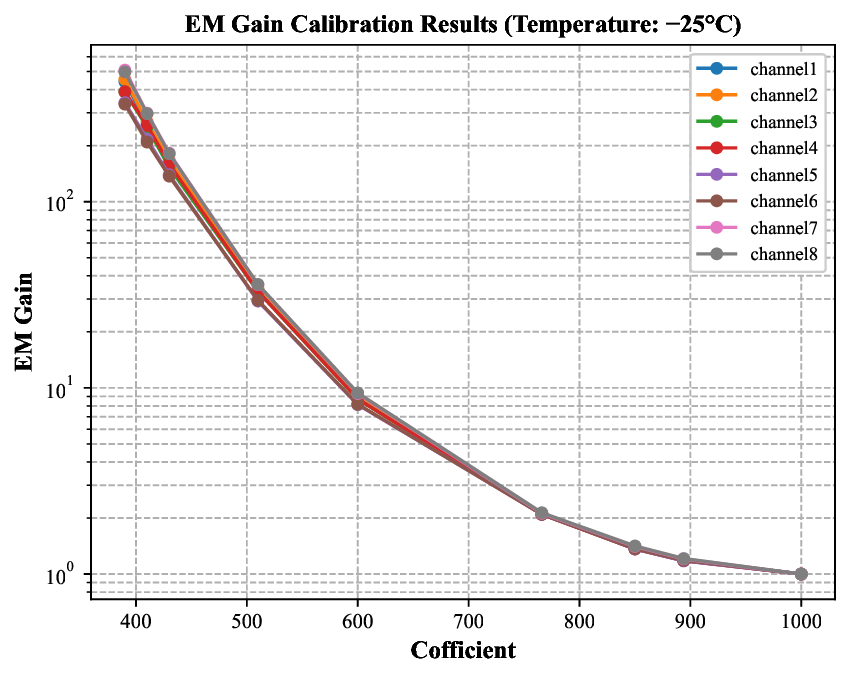}}

\caption{{Panels} (\textbf{a}--\textbf{c}) depict the EM Gain regulation curves as functions of injection coefficient under three cooling temperature conditions: $-$15 \textdegree C, $-$20 \textdegree C, and $-$25 \textdegree C, respectively, illustrating the calibration results.}
\label{fig7}
\end{figure}

In Equation (1), $\mathrm{flat}_{\mathrm{Gm}}$ and $\mathrm{bias}_{\mathrm{Gm}}$ denote the flat-field image and background image acquired under a specified EM Gain setting, respectively, while $\mathrm{exptime}_{\mathrm{Gm}}$ represents the corresponding exposure time. Conversely, $\mathrm{flat}_{\mathrm{1}}$ and $\mathrm{bias}_{\mathrm{1}}$ refer to the flat-field and background images captured at an EM Gain factor of 1$\times$, with $\mathrm{exptime}_{\mathrm{1}}$ indicating the exposure time applied in these acquisitions. Figure \ref{fig7} quantifies the EM Gain characteristics across three thermal operating regimes. At a fixed injection coefficient, the {$-$15~\textdegree C} condition fails to meet CPI-C operational thresholds with maximum EM Gain \textless 100$\times$, while {$-$25~\textdegree C} achieves sufficient EM Gain but exhibits degraded channel-to-channel uniformity (16.3\% deviation at maximum EM Gain). The {$-$20~\textdegree C} regime optimally satisfies both criteria, delivering the target gain range (1\textasciitilde150$\times$) with enhanced uniformity (9.35\% across channels, preliminarily compliant with CPI-C operational tolerances), thereby establishing this thermal condition as the baseline for engineering validation.

\subsection{Readout Noise}\label{sec32}
\begin{figure}
\centering
\includegraphics[width=9.5 cm]{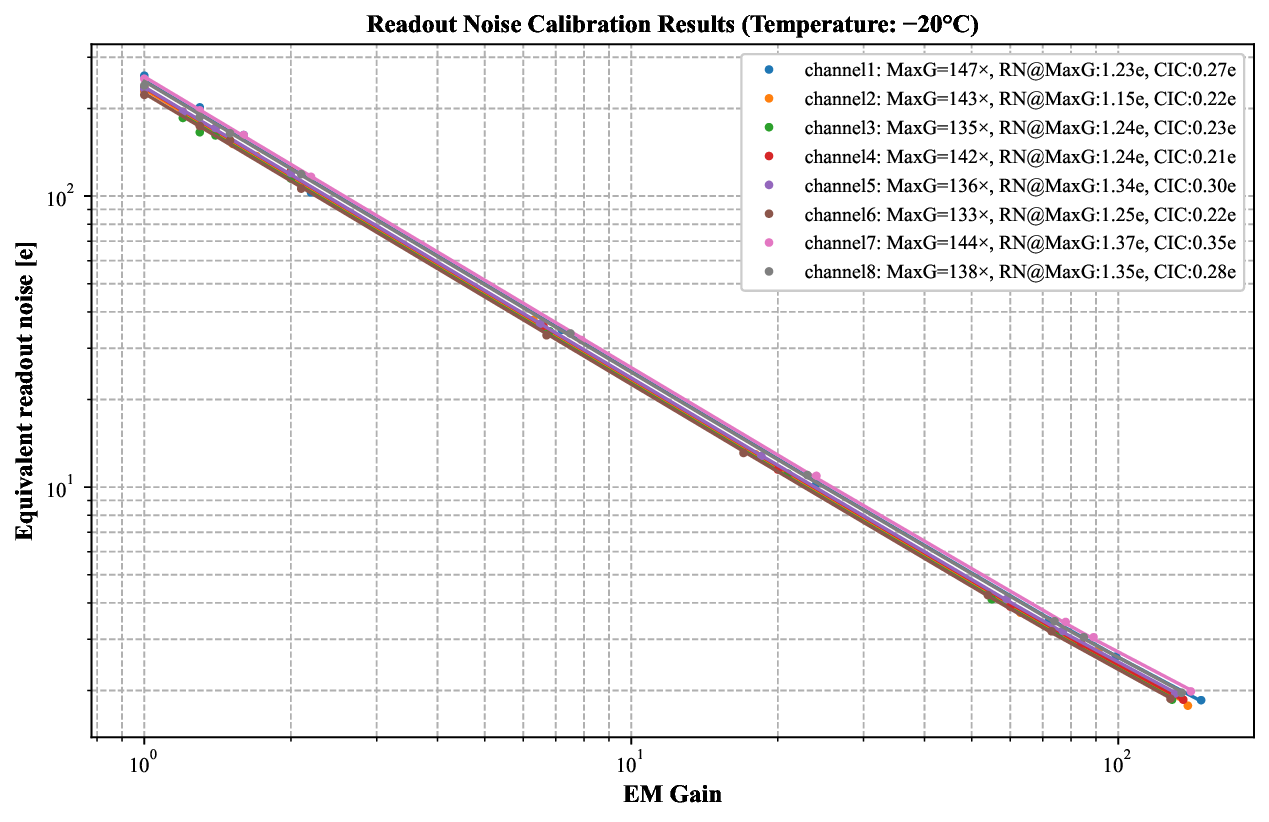}

\caption{Thereadout noise calibration results. As illustrated, the amplification of CIC noise is inadequate under the current operating conditions ($-$20 \textdegree C, Max EM Gain (MaxG) < 150$\times$).}\label{fig8}
 
\end{figure}

We characterized readout noise by systematically analyzing EM Gain calibration results at $-$20  \textdegree C. The quantitative relationship is mathematically formulated in \mbox{Equations (2) and (3),} where F represents noise factor \citep{Robbins2003}, and N denotes {the number of multiplication stages} in the EM Gain register. Furthermore, $\mathrm{RN}_{\mathrm{1\times}}$ signifies the readout noise at unity EM Gain, while CIC refers to clock-induced charge. Figure \ref{fig8} visualizes noise parameter dependencies on EM Gain.

\begin{equation}
\text{F}^\text{2} = 2 \times (\text{EM Gain} - \text{1})\times \text{EM Gain}^\frac{-\text{N+1}}{\text{N}}  + \frac{\text{1}}{\text{EM Gain}}
\end{equation}

\begin{equation}
\text{Equivalent RN}_\text{EM Gain} = \sqrt{{\text{F}^\text{2}} \times {\text{CIC}}  + \frac{\text{RN}^{\text{2}}_{\text{1$\times$}}}{\text{EM Gain}}}
\end{equation}

\subsection{Single-Photon Sensitivity}
The single-photon sensitivity, defined as the capability to distinguish individual photon events from noise, can be quantitatively evaluated through the {Noise-Equivalent Photon Count (NEPC)}. This metric arises from the fundamental requirement that the photon-induced signal power must exceed the total noise power in the system. Specifically, when the noise variance is less than the single-photon signal power, the condition NEPC < 1 ensures that photon-generated signals statistically dominate over noise fluctuations. The theoretical basis originates from the noise suppression and signal amplification model proposed by Robbins and Hadwen \citep{Robbins2003}, with experimental verification via the photon statistics based on Poisson distribution by Jedrkiewicz et al. \citep{Ottavia2011}. In this study, the NEPC of EMCCD detector is fundamentally constrained by the dynamic interplay between EM Gain and noise characteristics, as expressed in {Equation (4).}

\begin{equation}
\text{NEPC} = \frac{\sqrt{{\text{F}^\text{2}} \times {\text{CIC}}  + \frac{\text{RN}^{\text{2}}_{\text{1$\times$}}}{\text{EM Gain}}}}{\eta_{QE}\times\text{EM Gain}}
\end{equation}

\begin{figure}
\centering
\includegraphics[width=9.5 cm]{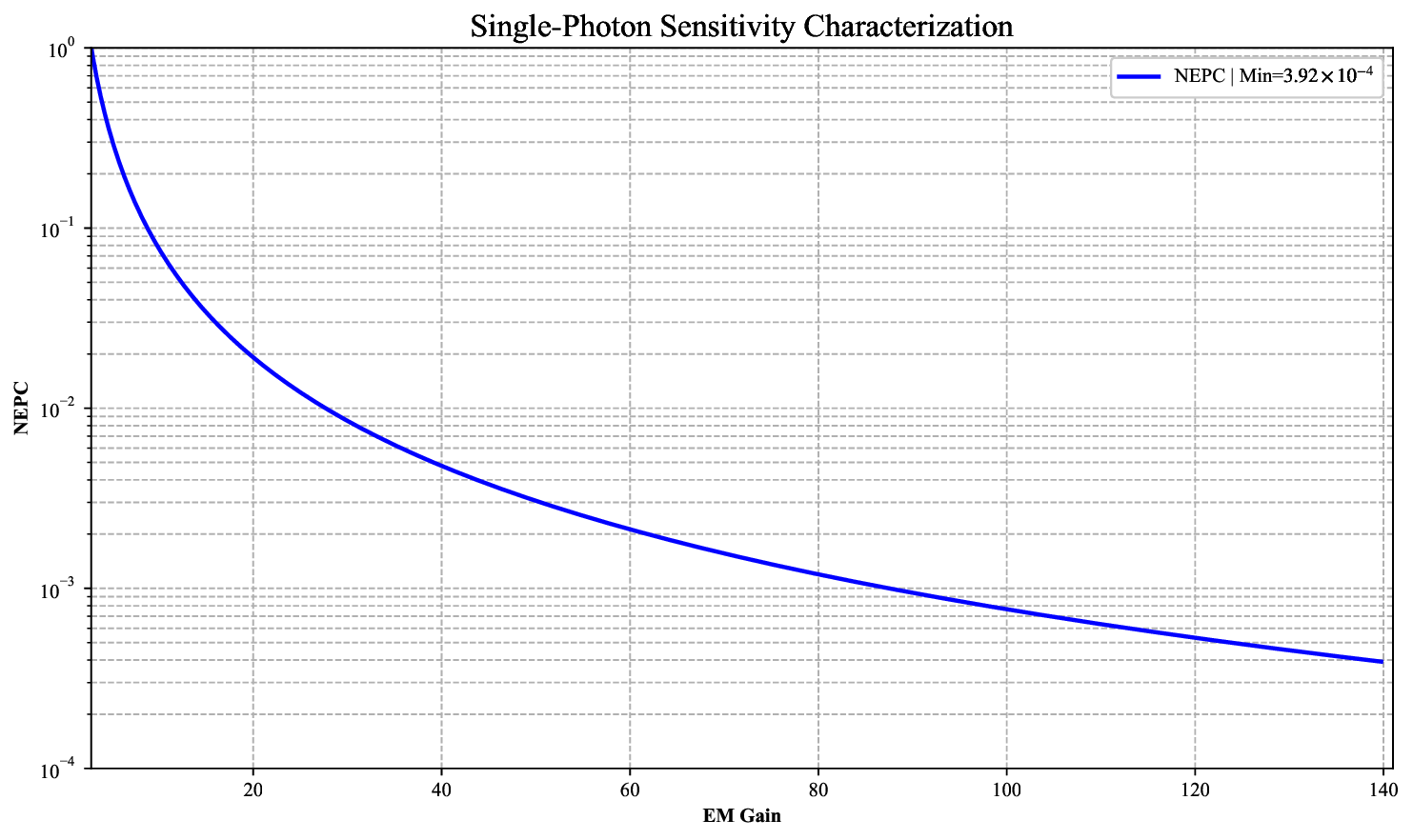}

\caption{{The} 
 single-photon sensitivity calibration result.\label{fig9}}
\end{figure}

Using calibrated EM Gain and noise characteristics at $-$20 \textdegree C, as established in \mbox{Sections \ref{sec31} and \ref{sec32},} we calibrated NEPC using quantum efficiency ($\eta_{QE}$ = 0.9) as the pivotal parameter. For multi-channel characterization, the input parameters were derived from the arithmetic mean of EM Gain and readout noise across all eight channels. As demonstrated in Figure \ref{fig9}, the NEPC exhibits a significant inverse correlation with EM Gain, reaching its minimum value of 3.92 $\times$ $10^{-4}$ under optimal amplification conditions. {This quantitative relationship explicitly illustrates the enhancement mechanism by which EM gain improves the detection efficiency of single photons, thereby validating the technical superiority of EMCCD in ultra-low-light detection scenarios.} The achieved NEPC magnitude highlights the critical role of electron multiplication in resolving single-photon events below the noise floor of standard imaging systems.

\unskip

\subsection{Inter-Channel Non-Uniformity Analysis}
\subsubsection{Theoretical Model}
EM Gain calibration results indicate significant {discrepancy} in inter-channel gain uniformity. This phenomenon can be attributed to the exacerbated non-uniformity in digital number (DN) values across channels during the amplification process. To quantitatively assess its impact on the CPI-C performance metric, we developed a DN-value non-uniformity metric evaluation model based on the characteristics of EM Gain-noise correlation. By constructing an inter-channel non-uniformity constraint function, this model offers a mathematical representation of the maximum allowable non-uniformity thresholds for various EM Gain settings. The non-uniformity of DN values across channels ($\text{DN}_{\text{CV}}$) and mathematical expression of the model are defined as follows:

\begin{equation}
\text{DN}_{\text{global}} = \frac{1}{N}\sum_{i=1}^{N} \text{DN}_{i}, \quad 
 \sigma_{\text{inter-channel}} = \sqrt{\sum_{i=1}^{N}(\text{DN}_{i} - \text{DN}_{\text{global}})^{2}}
\end{equation}

\begin{equation}
\text{DN}_{\text{CV}}=\frac{\sigma_{\text{inter-channel}}}{\text{DN}_{\text{global}}}\times 100\%
\end{equation}

\begin{equation}
\text{DN}_{\text{CV}}\le \sqrt{\frac{(\frac{2\pi}{100})^{2}\text{N}^{2}_{\text{photo}}-\text{RN}^{2}_{\text{EM Gain}}}{\text{N}_{\text{photo}}\text{EM Gain}^{2}}}
\end{equation}

\begin{figure}
\centering
\includegraphics[width=9.5 cm]{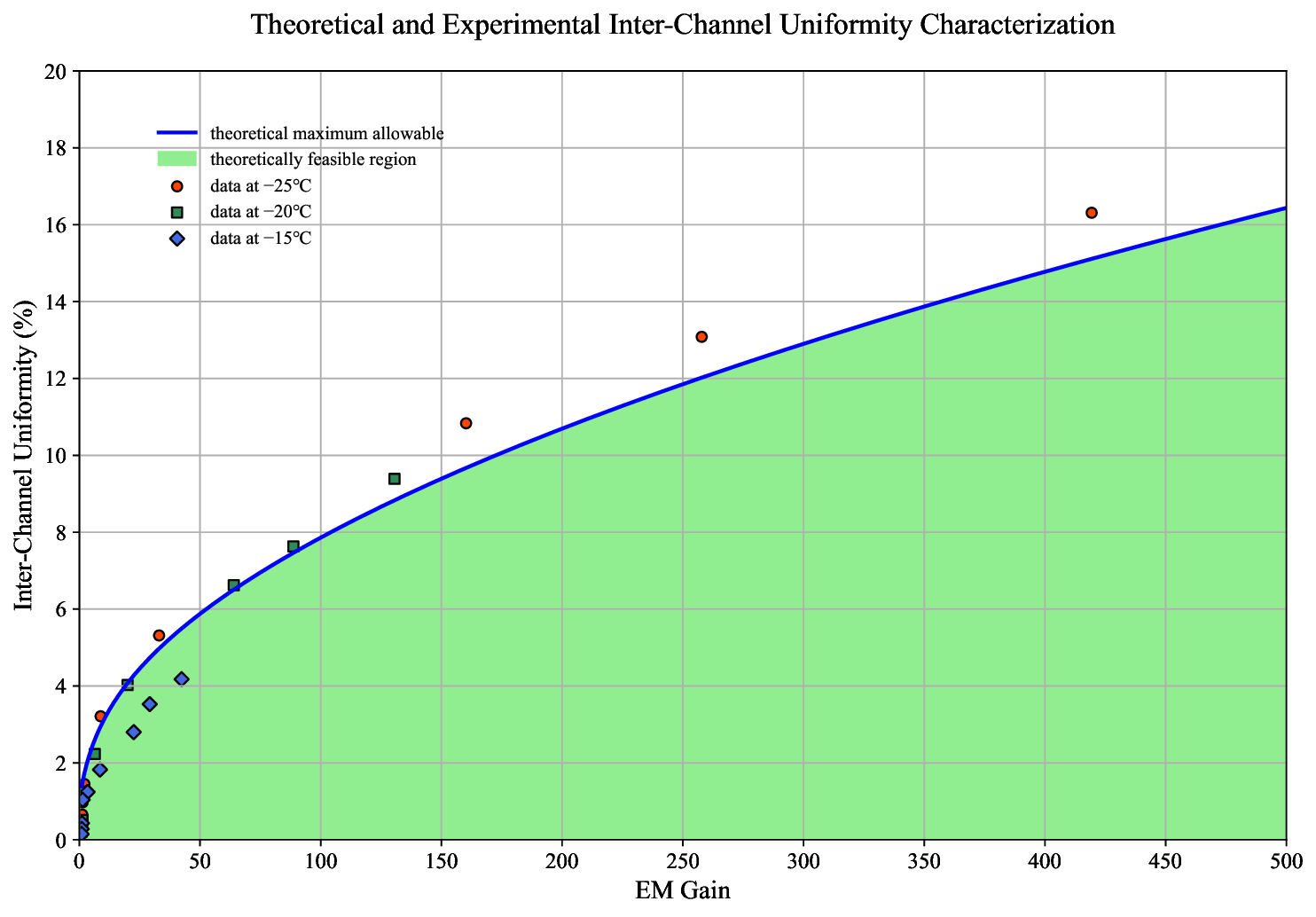}

\caption{{Theoretical}and experimental inter-channel non-uniformity characterization.\label{fig10}}
\end{figure}

According to the model established by Formulas (5)--(7), and in conjunction with the measurement results from the EM Gain calibration experiment detailed in Section \ref{sec31}, as well as the readout noise discussed in Section \ref{sec32}, we have {defined} a {allowable non-uniformity threshold boundary representing inter-channel non-uniformity} (as illustrated in \mbox{Figure \ref{fig10}}). Comparative analysis indicates that at a cooling temperature of $-$15 \textdegree C, the system operates within a low EM Gain range (30--50 times), and the uniformity of DN values across channels aligns {well} with theoretical expectations. However, when the cooling temperature is lowered to $-$20 \textdegree C, {localized non-compliance occurs} under maximum EM Gain conditions, nonetheless, {non-uniformity at other gain levels complies with} theoretical predictions for other EM Gain levels. In the deep cooling mode ($-$25 \textdegree C), there is a {pronounced} increase in EM Gain at identical injection voltages, which leads to significant nonlinear amplification of channel responses. This causes DN non-uniformity deviations exceeding theoretical bounds by 3--5\%. The nonlinear coupling effects resulting from gain jumps and readout noise {result in exponentially diverging response distortion} within each channel's transmission function at high gain levels, thereby impacting overall system response consistency.

To address this issue, this paper proposes a systematic solution. Given the limitations imposed by existing detector hardware architecture regarding adjustability, we introduce a robust correction {algorithm} designed to achieve dynamic compensation for {excessive non-uniformity} deviations {within the digital image processing pipeline}.

\subsubsection{Robust Correction {Algorithm}}
\begin{figure}

\centering
\subfloat[\centering]{\includegraphics[width=3.5cm]{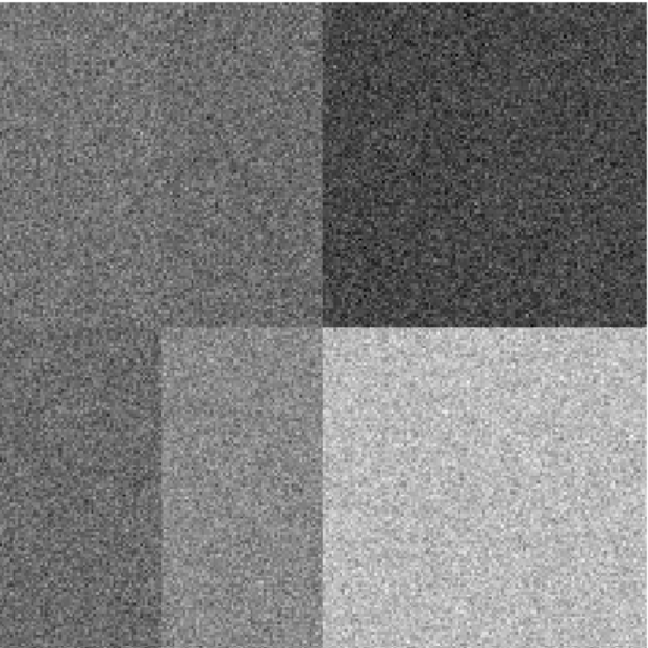}}
\hfill
\subfloat[\centering]{\includegraphics[width=3.5cm]{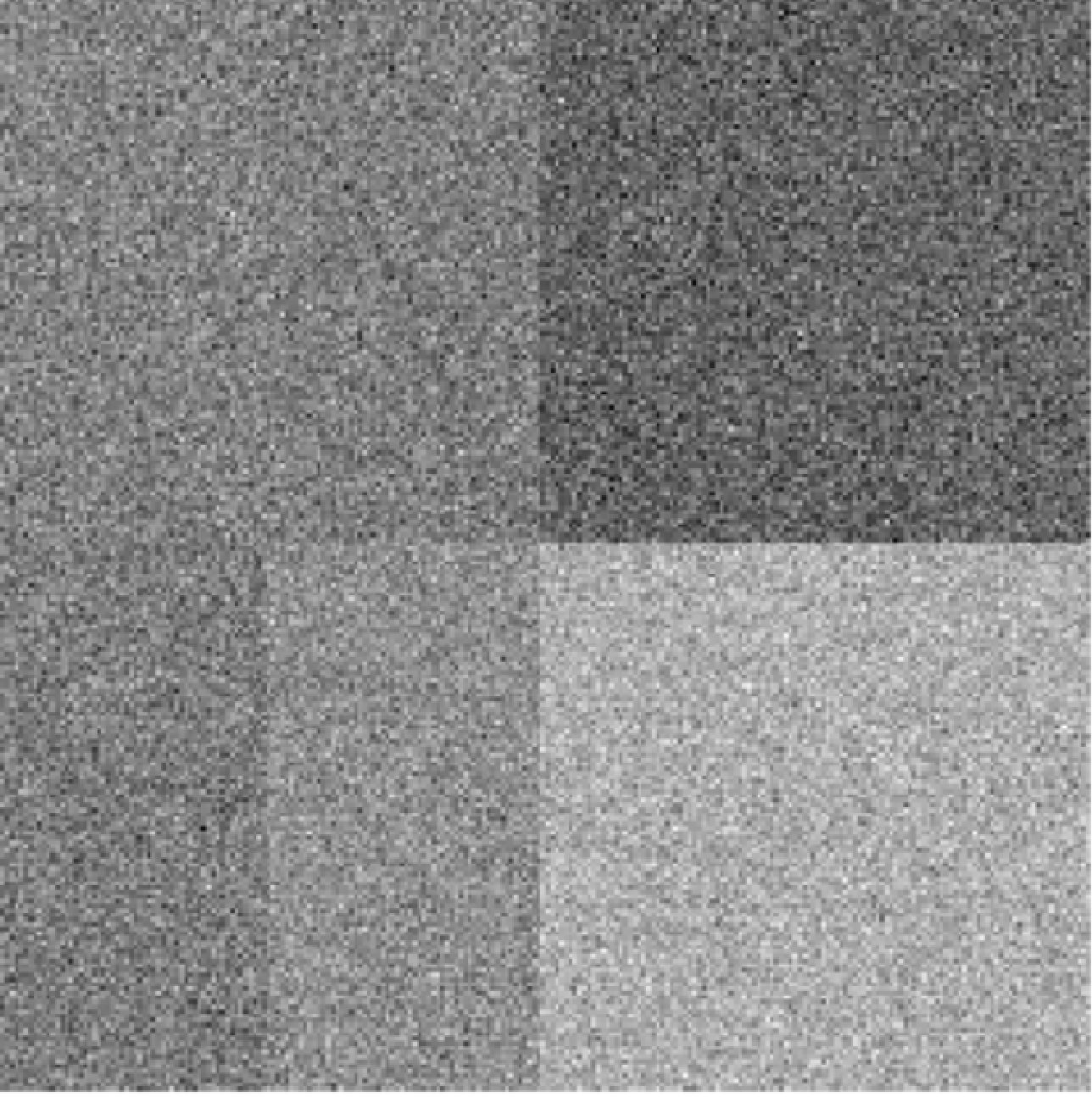}}
\hfill
\subfloat[\centering]{\includegraphics[width=3.5cm]{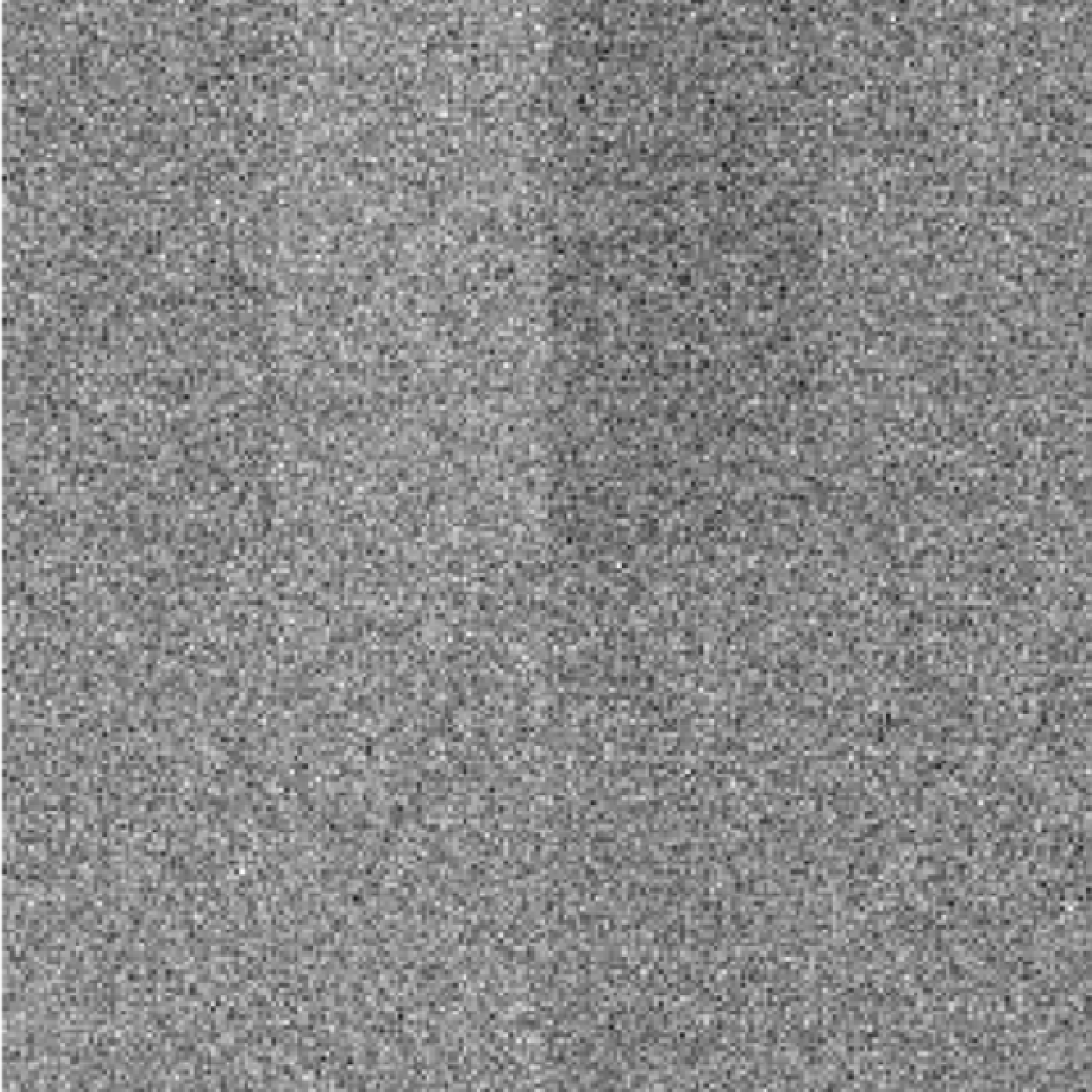}}\\
\hspace{4cm}
\subfloat[\centering]{\includegraphics[width=14cm]{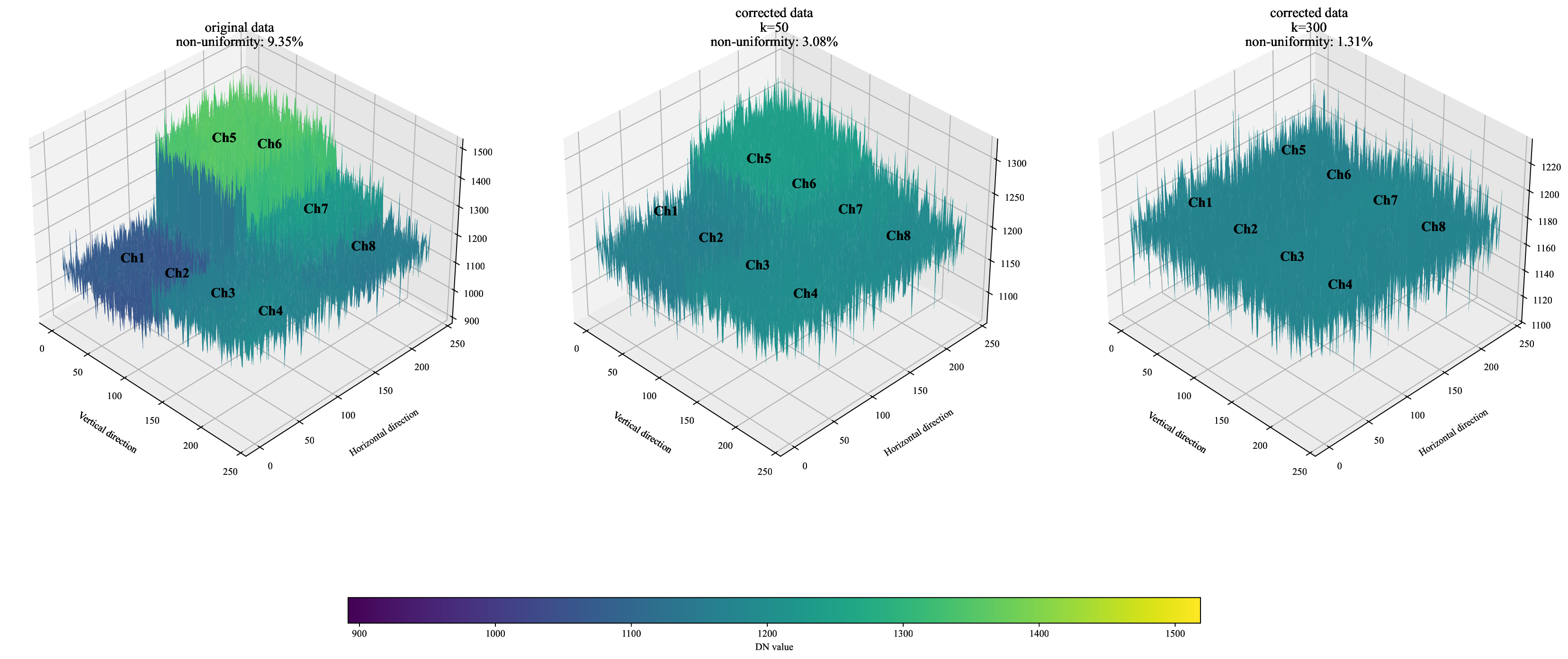}}

\caption{The systematic evaluation of {algorithm} effectiveness. (\textbf{a}) displays the original flat-field response distributions following background correction. Panels (\textbf{b},\textbf{c}) illustrate the corrected results after {algorithm} processing, while (\textbf{d}) presents the distribution profiles of channel DN values. Notably, as evidenced in the comparative diagrams (\textbf{a}--\textbf{c}), Channel 1 (Ch1) maintains consistent spatial positioning in the upper-right quadrant of each schematic layout.
} \label{fig11}
\end{figure}

We propose a robust non-uniformity correction {algorithm} using dynamic response compensation (Formulas (8) and (9)), which enables adaptive nonlinear correction across channels. The {algorithm} operates under conditions of $-$20 \textdegree C with MaxG, validated using 1000 bias frames and 1000 flat-field image frames. During preprocessing, {bias-subtracted} flat-field images are organized into a three-dimensional data cube of dimensions 1000 (T) $\times$ 240 (H) $\times$ 240 (V). A master flat-field image is generated by computing the temporal mean along the time axis (T dimension), as illustrated in Figure \ref{fig11}. 

\begin{equation}
\text{DN}_{\text{corr}}^{(k)} = \text{DN}_{\text{origin}} \odot \mathbf{g}^{(k)}
\end{equation}

\begin{equation}
\mathbf{g}^{(k+1)} = \mathbf{g}^{(k)} \odot 
\begin{cases} 
g_{\min}^{(k)}, & \left( \frac{c^{(k)}}{\mu(c^{(k)})} \right)^{\alpha^{(k)}} \le g_{\min}^{(k)} \\[6pt]
\left( \frac{c^{(k)}}{\mu(c^{(k)})} \right)^{\alpha^{(k)}}, & g_{\min}^{(k)} < \left( \frac{c^{(k)}}{\mu(c^{(k)})} \right)^{\alpha^{(k)}} < g_{\max}^{(k)} \\[6pt]
g_{\max}^{(k)}, & \left( \frac{c^{(k)}}{\mu(c^{(k)})} \right)^{\alpha^{(k)}} \ge g_{\max}^{(k)}
\end{cases}
\end{equation}

The {algorithm} optimizes inter-channel uniformity via dynamic gain adjustment. Key components include the following: original DN values ({$\text{DN}_{\text{origin}}$}) as input data, ${k}$ denotes the iteration index, iterative gain vector ($\mathbf{g}^{(k)}$) for channel-wise correction generating corrected DN values ($\text{DN}_{\text{corr}}^{(k)}$), corrected mean ($\mu(c^{(k)})$) evaluating uniformity, dynamic smoothing factor ($\alpha^{(k)}$) regulating adjustment sensitivity ($\alpha^{(k)} < 1$ suppresses oscillations, $\alpha^{(k)} > 1$ accelerates convergence), and dynamic gain bounds ($g_{\min}^{(k)}$, $g_{\max}^{(k)}$) constraining stepwise adjustments via {saturation clipping} to prevent overshooting. Parameters were selected based on initial gain $\mathbf{g}^{(0)} = 1$ preserves data integrity, $\alpha^{(k)}$ adapts to uniformity improvement rates, gain bounds ($g_{\min}^{(k)} = \frac{\text{DN}^{(k)}_{\text{mean}}}{\text{DN}^{(k)}_{\text{max}}}$, $g_{\max}^{(k)}= \frac{\text{DN}^{(k)}_{\text{mean}}}{\text{DN}^{(k)}_{\text{min}}}$). 

Experimental validation confirms the {algorithm}'s efficacy in suppressing inter-channel non-uniformity. Through iterative calibration of per-channel DN values, the {algorithm} {iteratively suppresses non-uniformity, driving channel DN values toward a unified mean}. The results reveal a trade-off between correction accuracy and real-time performance, modulated by the iteration count k. Increasing k progressively enhances correction: at low k, channel DN values achieve satisfactory consistency, at high k, uniformity approaches near-ideal levels. However, higher k incurs increased computational load, {proportionally increasing} processing time. Specifically, low k prioritizes rapid response, while high k emphasizes accuracy at the expense of latency.

\section{Discussion}\label{sec4}

This study systematically {addresses} the performance calibration and optimization of the WFS's EMCCD detector for the CPI-C onboard the CSST, {establishing critical} technical validation for space-based high-contrast imaging. {Through comprehensive characterization spanning EMCCD chip screening, EM Gain calibration, noise modeling, and inter-channel non-uniformity mitigation, we established key parameters including EM Gain range and readout noise.} The results demonstrate that multi-criteria chip selection (e.g., balancing noise and resolution) and temperature-voltage co-optimization (e.g., achieving minimal EM Gain deviation of 9.35\%  at $-$20 \textdegree C) significantly enhance the sensor's signal-to-noise ratio (SNR) and reliability. 

EM Gain and readout noise optimization proved essential for low-noise detection. Experiments {demonstrated} a nonlinear increase in EM Gain as the temperature decreased. {At $-$20 \textdegree C, an inter-channel deviation of no more than 9.35\% was observed, underscoring the importance of precise thermal management.} However, {CIC noise became the dominant limitation when EM Gain exceeds} 150$\times$, necessitating future improvements in chip fabrication or algorithmic compensation to achieve sub-electron noise levels.

The EMCCD chip was selected for detector fabrication due to its multi-channel readout capability and high frame rate, enabling rapid wavefront capture. A key challenge lies in ensuring {clock voltage uniformity} across channels. Under current circuit conditions, observed DN-value non-uniformity exceeds expectations at high EM Gains. We therefore propose a correction {algorithm} that dynamically adjusts gain coefficient $\alpha$ and boundary constraints ($g_{\min}^{(k)}$, $g_{\max}^{(k)}$). This approach balances convergence efficiency and accuracy: initial iterations use larger $\alpha$ (0.8) and {relaxed} gain bounds (0.5–1.5) for rapid error reduction, while later stages tighten parameters ($\alpha$ = 0.3, bounds 0.8–1.2) to prevent over-correction. Experimental data confirm progressive convergence toward ideal correction values. However, real-time applications face theoretical limits with software-only optimization, underscoring the need for integrated hardware-circuit improvements. {Specifically, the replacement of a single register clock driver (RCD) that controls four channels with four independent RCDs would improve voltage consistency across all channels.} Combined with algorithmic correction, this hardware-software co-design ensures optimal DN-value uniformity. These insights inform detector circuit design and establish foundations for future system integration. Optical component calibration will be detailed separately.

While ground-based tests {confirm} sensor performance, {spaceflight qualification demands additional evaluation.}  {Uncharacterized risk factors} include in-orbit radiation impacts on EMCCD longevity and noise, necessitating long-term stability tests. Furthermore, {ground-based static microlens testing does not fully replicate} dynamic observation conditions; {future studies should integrate} simulated dynamic wavefront disturbances to verify real-time closed-loop correction. Compared to existing spaceborne sensors (e.g., Hubble's Fine Guidance Sensor), the current design demonstrates advantages in noise suppression and frame rate but requires further {alignment with spacecraft resource limitations} to meet space payload requirements.

This study {proposes a comprehensive technological roadmap} for advancing space-based high-contrast imaging. First, EMCCD-based WFS can be {deployed in next-generation exoplanet missions} (e.g., exoplanet spectroscopy), with multi-band calibration enhancing sensitivity. Second, {the integration of advanced adaptive optics algorithms} (e.g., deep learning-based wavefront prediction) may {surpass} current contrast limits, {enabling the detection of fainter exoplanets.} Finally, {the operationalization of the CSST will leverage the calibration methodologies and experimental data presented in this study for in-orbit calibration and fault diagnosis, further solidifying China's leadership in exoplanet exploration.} Future efforts will focus on in-orbit validation, radiation-hardened design, and CPI-C subsystem integration, {effectively bridging theoretical breakthroughs with practical spacecraft engineering implementation}.

\section{Conclusions}\label{sec5}
This study {presents} a {comprehensive} characterization of critical performance parameters and calibration data for the custom EMCCD detector integrated into the CPI-C's WFS. Experimental validation confirms that the detector {satisfies} CPI-C's initial specifications, thereby establishing its {mission-readiness}. However, the current detector performance {does not yet meet space-grade requirements}. Key metrics, such as readout noise suppression and inter-channel uniformity, require further optimization. Additionally, {precision recalibration of critical parameters}—including noise characteristics and response linearity under cryogenic conditions—will be conducted in subsequent verification phases. QM calibration data provide {essential} insights for refining the design and manufacturing protocols of the FM. Collectively, these results establish a robust framework for developing the flight-ready sensor. {By integrating the insights gained from this phase}, future iterations of the WFS EMCCD will achieve enhanced performance, featuring sub-electron readout noise and compatibility with real-time adaptive optics. This advancement will significantly {enhance} CPI-C’s precision and reliability in {achieving} its scientific objectives.

\bibliographystyle{unsrtnat}
\bibliography{references}  






\end{document}